\documentclass[useAMS,usenatbib]{mn2e}
\usepackage{graphicx}
\usepackage{amssymb}
\usepackage{color}

\title[Nonthermal Emission from W51C]
{On the Nonthermal Emission from the Supernova Remnant W51C}
\author[Fang and Zhang]{Jun Fang , Li Zhang \thanks{lizhang@ynu.edu.cn} \\
Department of Physics, Yunnan University, Kunming, China}

\begin{document}
\pagerange{\pageref{firstpage}--\pageref{lastpage}} \pubyear{2008}

\maketitle

\label{firstpage}

\begin{abstract}
The middle-aged supernova remnant (SNR) W51C is an interesting
source for the interaction of the shell with a molecular cloud. The
shell emits intense radio synchrotron photons, and high-energy
gamma-rays from the remnant have been detected using the {\it Fermi}
Large Area Telescope (LAT), the H.E.S.S. telescope, and the Milagro
gamma-ray observatory. Based on a semi-analytical approach to the
nonlinear shock acceleration process, we investigate the multiband
nonthermal emission from W51C. The result shows that the radio
emission from the remnant can be explained as synchrotron radiation
of the electrons accelerated by a part of the shock flowing into the
ambient medium. On the other hand, the high-energy gamma-rays
detected by the {\it Fermi} LAT are mainly produced via
proton-proton collisions of the high-energy protons with the ambient
matter in the molecular cloud overtaken by the other part of the
shock. We propose a possible explanation of the multiband nonthermal
emission from W51C, and it can be concluded that a molecular cloud
overtaken by a shock wave can be an important emitter in GeV
$\gamma$-rays.

\end{abstract}

\section{Introduction}
\label{sec:intro}

SNRs are broadly thought as the primary accelerators of the Galactic
cosmic rays accelerated through diffusive shock acceleration in the
shock waves \citep[e.g.,][]{BE87,BV07}. Observations of SNRs in the
high-energy $\gamma$-ray band provide evidence for acceleration of
particles to relativistic energies \citep[e.g.,][]{Aea08c}. However,
both electrons and nuclei can be accelerated to relativistic
energies, and the high-energy $\gamma$-rays from SNRs can usually be
explained either as proton-proton (p-p) interaction between the
relativistic protons and the ambient matter or as inverse Compton
scattering of the electrons on the soft photons
\citep[e.g.,][]{Yea06,Aea07,BV08,Fea08,Fea09,ZA10}. The
interpretation of the $\gamma$-rays produced via the inverse Compton
scattering makes the evidence of the argument that the nuclei in the
cosmic rays are accelerated by the SNR shells indirectly.

P-p collisions can be greatly enhanced if a SNR shell interacts with
a molecular cloud, and observations of $\gamma$-rays from this
system provide evidence of the nuclei in the cosmic rays being
accelerated by SNRs \cite[e.g.,][]{Aea94}. Moreover, $\gamma$-rays
from such systems have probably been detected with H.E.S.S.
\citep[e.g.,][]{Aea08a,Aea08b} and EGRET \citep[][]{Hea99}.

The SNR W51C appears as a partial shell of $\sim30'$ diameter in the
radio continuum with an open northern part \citep[][]{MK94}. From
the observations of the two 1720 MHz OH masers toward the north of
the shell \citep[][]{Gea97} and the detection of shocked atomic and
molecular gases in this direction \citep[][]{KM97a,KM97b}, it is
argued that this remnant is interacting with a molecular cloud.

Both shell-type and center-filled morphologies were indicated in the
X-ray observations with ROSAT, and an age of $\sim3\times10^4$ yr
and an explosion energy $\sim3.6\times10^{51}$ erg were estimated
base on either the Sedov or the evaporative models
\citep[][]{Kea95}. An extended source, CXO J192318.5+1403035,
composed of a relatively bright core surrounded by a diffuse
envelope was discovered with ASCA \citep[][]{Kea02}, and it was also
detected with Chandra \citep[][]{Kea05}. The core-envelope structure
and its spectral properties suggest that the source is a pulsar wind
nebula (PWN) associated with the SNR 51C, and a putative pulsar is
argued to exist inside it.

In TeV $\gamma$-rays, an extended source HESS J1923+141 coincident
with W51C have been found using H.E.S.S. \citep[][]{Fiea09}.
Moreover, \citet{Abea09a} recently reported a possible excess of
multi-TeV gamma-rays towards W51C with a flux of
$39.4\pm11.5\times10^{-17}$ TeV$^{-1}$ cm$^{-2}$ s$^{-1}$ at 35 TeV
from the Milagro observation. Furthermore, the GeV $\gamma$-rays
from the SNR, which is known to be interacting with a molecular
cloud, with a luminosity of $\sim1\times10^{36}$ erg s$^{-1}$ for a
distance of 6 kpc, have been detected with the LAT on board the {\it
Fermi} Gamma-ray Space Telescope \citep[][]{Abea09b}. The
$\gamma$-ray spectrum cannot be fitted with a single power law and
steepens above a few GeV. \citet[][]{Abea09b} discussed the
nonthermal radiative properties of the radio emission and
$\gamma$-rays from the SNR using a broken power-law with the same
index for the momentum distribution of the radiating
electrons/protons. Their result shows that p-p interaction is more
likely the main process to produce the observed $\gamma$-rays since
the observed radio synchrotron spectrum cannot be well reproduced in
the bremsstrahlung-dominated case and an origin of inverse Compton
scattering for the $\gamma$-rays needs a low density of hydrogen
($<0.1$ cm$^{-1}$) in the SNR, which conflicts with the constraints
from observations in the radio and X-ray bands.

Semi-analytical methods to the nonlinear diffusive shock
acceleration process have been widely used to investigate the
nonthermal radiative properties of SNRs
\citep[e.g.,][]{Mea09,Fea09,Eea10}. In this paper, we study the
nonthermal radiative properties of the SNR W51C based on a
semi-analytical method to the nonlinear diffusive shock acceleration
mechanism. In our model, a part of the SNR shell transports into the
molecular cloud (MC), then the shock is greatly decelerated due to
the large density of matter in the MC. As a result, the spectra of
the accelerated protons and electrons are determined in a case with
a low Mach number. We find that the observed $\gamma$-ray spectrum
for W51C can be well reproduced via p-p collisions of the
accelerated protons in the part of shell interacting with the MC,
and the radio emission from the SNR can be explained as the
radiation of the electrons accelerated by the other part of the SNR
shock freely expanding into the ambient interstellar space.

\section{Model and Results}
\label{sec:model}

The pitch-angle averaged steady-state distribution of the protons
accelerated at a shock in one dimension satisfies the diffusive
transport equation \citep[e.g.,][]{MD01, B02, ABG08},
\begin{eqnarray}
\nonumber \frac{\partial}{\partial x}\left
[D\frac{\partial}{\partial x}f(x, p)\right ] &-& u\frac{\partial
f(x, p)}{\partial x} \\ &+& \frac{1}{3}\frac{du}{dx}p\frac{\partial
f(x, p)}{\partial p} + Q(x, p) = 0, \label{eq:diff}
\end{eqnarray}
where the coordinate $x$ is directed along the shock normal from
downstream to upstream, $D$ is the diffusion coefficient and $u$ is
the fluid velocity in the shock frame, which equals $u_2$ downstream
($x<0$) and changes continuously upstream, from $u_1$ immediately
upstream ($x=0^+$) of the subshock to $u_0$ at upstream infinity
($x=+\infty$).  For the Bohm diffusion, $D=pc^2/(3eB)$, where $B$ is
the local magnetic field strength. With the assumption that the
particles are injected at immediate upstream of the subshock, the
source function can be written as $Q(x, p)=Q_0(p)\delta(x)$. For
monoenergetic injection, $Q_0(p)$ is
\begin{equation}
Q_0(p) = \frac{\eta n_{\rm{gas, }1}u_1}{4\pi
p_{\rm{inj}}^2}\delta(p-p_{\rm{inj}})\;\; , \label{eq:Q0}
\end{equation}
where $p_{\rm{inj}}$ is the injection momentum, $n_{\rm{gas, }1}$ is
the gas density at $x=0^+$ and $\eta$ is the fraction of particles
injected in the acceleration process. With the injection recipe
known as thermal leakage, $\eta$ can be described as $\eta =
4(R_{\rm{sub}}-1)\xi^3e^{-\xi^2}/3\pi^{1/2}$ \citep{BGV05, ABG08},
where $R_{\rm{sub}}=u_1/u_2$ is the compression factor at the
subshock and $\xi$ is a parameter of the order of 2--4 describing
the injection momentum of the thermal particles in the downstream
region ($p_{\rm{inj}}=\xi p_{\rm{th,}2}$). We use $\xi=3.5$ as in
\citet{ABG08}, $p_{\rm{th,}2}=(2m_pk_{\rm{B}}T_2)^{1/2}$ is the
thermal peak momentum of the particles in the downstream fluid with
temperature $T_2$, $m_p$ is the proton mass and $k_{\rm{B}}$ is the
Boltzmann constant. Assuming the heating of the gas upstream is
adiabatic, with the conservation condition of momentum fluxes
between the two sides of the subshock, we can derive the relation
between the temperature of the gas far upstream $T_0$ and $T_2$,
i.e., $T_2 = (\gamma_g M_0^2/R_{\rm tot})[R_{\rm sub}/R_{\rm tot} -
1/R_{\rm tot} + (1/\gamma_g M_0^2)(R_{\rm tot}/R_{\rm
sub})^{\gamma_g}]T_0$, where $M_0$ is the fluid Mach number far
upstream, $R_{\rm tot}=u_0/u_2$ is the total compression factor,
$\gamma_g$ is the ratio of specific heats ($\gamma_g=5/3$ for an
ideal gas)

Particles with larger momenta move farther away from the shock than
those with lower momenta, only particles with momentum $\geq p$ can
reach the point $x_p$ \citep[see details in][]{B02}, and thus the
pressure of the accelerated particles at the point can be described
as
\begin{equation}
P_{{\rm{CR}}, p}=\frac{4\pi}{3}\int^{p_{\rm max}}_{p}dp' p'^3 v(p')
f_0(p') \;\; , \label{PCR}
\end{equation}
where $v(p)$ is the velocity of particles with momentum $p$.
Furthermore, the particle distribution function $f_0(p)$ at the
shock can be implicitly written as \citep{B02}
\begin{eqnarray}
\nonumber f_0(p) & = & \left[\frac{3R_{\rm{tot}}}{R_{\rm{tot}}U(p) -
1} \right]\frac{\eta n_{\rm ism }}{4\pi p_{\rm{inj}}^3}
\\ && \times \exp\left[ -
\int^{p_{\rm{max}}}_{p_{\rm{inj}}}\frac{dp'}{p'}\frac{3R_{\rm
tot}U(p')}{R_{\rm{tot}}U(p') - 1}\right], \label{eq:f0}
\end{eqnarray}
where $n_{\rm ism}$ is the gas density far upstream ($x=+\infty$),
$p_{\rm max}$ is the maximum momentum of the accelerated protons,
and it can be estimated as \citep{LC83,AA99}
\begin{equation}
p_{\rm max} = 5.3\times10^4\frac{B}{100 \mu\rm{G}}
t_3\left(\frac{u_0}{10^8 {\rm cm\ s^{-1}}}\right)^2 \;\;  m_p c ,
\label{eq:Pmax}
\end{equation}
for a "Bohm limit" according to the standard diffusive shock
acceleration theory, where $B$ the magnetic field strength, $t_3$ is
the age of the host SNR in units of $10^3$ yr. $U(p)$ can be solved
using an equation deduced from the conservation of the mass and
momentum fluxes with the boundary condition $U(p_{\rm inj}) = R_{\rm
sub}/R_{\rm tot}$ and $U(p_{\rm max})=1$, and then a value of
$R_{\rm sub}$ can be achieved by an iterative procedure to satisfy
the boundary conditions \citep{B02}.

The electrons have the same spectrum of the protons up to a maximum
energy determined by synchrotron losses, and the spectrum can be
described as \citep[e.g.,][]{Fea09}
\begin{equation}
f_e(x, p)=K_{ep}f(x, p)\exp(-E(p)/E_{{\rm max}, e}), \label{fe}
\end{equation}
where $E(p)$ is the kinetic energy of the electrons, and the
electron/proton ratio $K_{ep}$ is treated as a parameter. By
equating the synchrotron loss time with the acceleration time, the
cutoff energy $E_{{\rm max}, e}$ can be estimated as
\citep[][]{Bet02}
\begin{equation}
\label{eq:Emaxe} E_{{\rm max}, e} = 6\times10^7
\left(\frac{u_0}{10^8 {\rm cm\ s^{-1}}}\right)\left[\frac{R_{\rm
tot}-1}{R_{\rm tot}(1+R_{\rm sub}R_{\rm tot})}\left(\frac{10\mu {\rm
G}}{B}\right)\right]^{1/2} {\rm MeV} .
\end{equation}

Assuming the accelerated particles distribute homogeneously and most
of the emission is from downstream of the shock, and using the
distribution function at the shock to represent the particle
distribution in the whole emitting zone, the volume-averaged
emissivity for photons produced via p-p interaction can be written
as
\begin{equation}
Q(E)=4\pi n_{\rm gas}\int dE_{\rm p} J_{\rm p}(E_{\rm
p})\frac{d\sigma(E, E_{\rm p})}{dE} \;\; , \label{eq:PP}
\end{equation}
where $E_{\rm p}$ is the proton kinetic energy, $n_{\rm gas}$ is the
ambient gas number density, and $J_{\rm p}(E_{\rm p})=v p^2 f_0(p)
dp/dE_{\rm p}$ is the volume-averaged proton density and $v$ is the
particles' velocity. We use the differential cross-section for
photons $d\sigma(E, E_{\rm p})/dE$ presented in \citet{Ka06} to
calculate the hadronic $\gamma$-rays produced via p-p collisions.
Finally, the photon flux observed at the earth can be obtained with
\begin{equation}
F(E)=\frac{VQ(E)}{4\pi d^2} \;\; , \label{eq:Flux}
\end{equation}
where $d$ is the distance from the earth to the source and $V$ is
the average emitting volume of the source and can be estimated by
$V\approx(4\pi/3)R_{\rm snr}^3 /R_{\rm tot}$ \citep{E00}, here
$R_{\rm snr}$ is the radius of the SNR.

\begin{figure}
\includegraphics[scale=1.2]{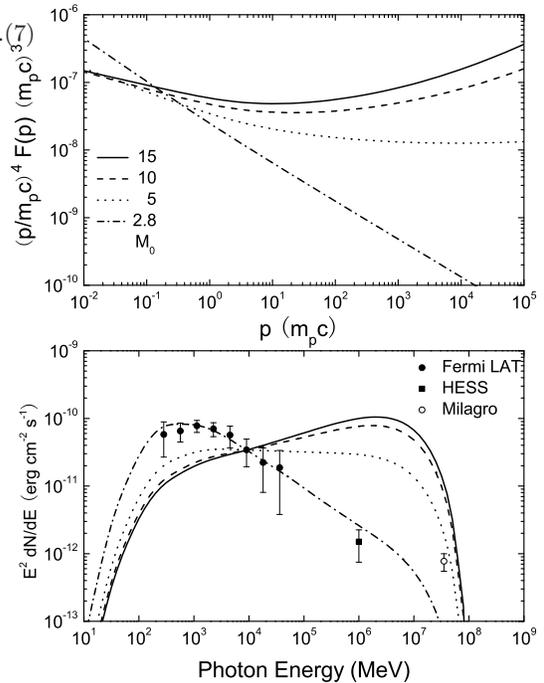}
\caption{\label{figdm} The resulting proton spectra (upper panel)
and the spectral energy distribution of photons from the p-p
collisions (lower panel) for $M_0=15$, $n_{\rm ism}=0.34$ cm$^{-3}$
(solid line); $M_0=10$, $n_{\rm ism}=0.53$ cm$^{-3}$ (dashed line);
$M_0=5$, $n_{\rm ism}=1.51$ cm$^{-3}$ (dotted line); and $M_0=2.8$,
$n_{\rm ism}=14.4$ cm$^{-3}$ (dash-dotted line). The other
parameters are $d = 6$ kpc, $R_{\rm snr}=30$ pc, $T_{\rm
up}=0.5\times10^6$ K, and $p_{\rm max}=10^5$ m$_{\rm p}$c. For each
$M_0$, $n_{\rm ism}$ is normalized to make the resulting flux at 9
GeV consist with the one observed by the {\it Fermi} LAT. The flux
points for the SNR W51C obtained with the {\it Fermi} LAT
\citep[][]{Abea09b}, H.E.S.S. \citep[][]{Fiea09}, and Milagro
\citep[][]{Abea09a} are shown in the figure for comparison.}
\end{figure}

The SNR W51C has an age of $\sim3\times10^4$ yr \citep{Kea05} and a
radius of $\sim30$ pc for a distance of 6 kpc \citep{MK94}. With $d
= 6$ kpc, $R_{\rm snr}=30$ pc, $T_{\rm up}=0.5\times10^6$ K, and
$p_{\rm max}=10^5$ m$_p$c, the resulting proton spectra and the
spectral energy distribution of the p-p collisions for the protons
with different Mach number are illustrated in Fig.\ref{figdm}. The
results are weakly sensitive to the temperature upstream of the
shock, and a temperature of $0.5\times10^6$ K is used in the
calculation, which corresponds to a sound speed of 83 km s$^{-1}$.
The observed results with the {\it Fermi} LAT \citep[][]{Abea09b},
H.E.S.S. \citep[][]{Fiea09}, and Milagro \citep[][]{Abea09a} are
shown in the figure for comparison. For each $M_0$, $n_{\rm ism}$ is
normalized to make the resulting flux at 9 GeV consistent with the
one observed with the {\it Fermi} LAT. We find that the spectrum
given by the {\it Fermi} LAT can only be reproduced with a low Mach
number $\sim3$ with those parameters. For such a low Mach number,
the reaction of the accelerated particles on the shock is
negligible, and the value of $R_{\rm sub}$ deviates little from that
of $R_{\rm tot}$. In fact, for $M_0=2.8$, $R_{\rm sub}$ and $R_{\rm
tot}$ are 2.83 and 2.91, respectively. The index for the momentum
distribution of the accelerated protons/electrons can be estimated
as $3(\gamma_g+1)M_0^2/(2(M_0^2-1))$, which is equal to 4.5 for
$M_0=3$ and $\gamma_g=5/3$. With such an index, the observed radio
emission cannot be well explained with the synchrotron emission of
the electrons, although the emission observed with the {\it Fermi}
LAT in the $\gamma$-ray band can be well reproduced as the p-p
interaction of the accelerated protons.

\begin{figure}
\includegraphics[scale=1.2]{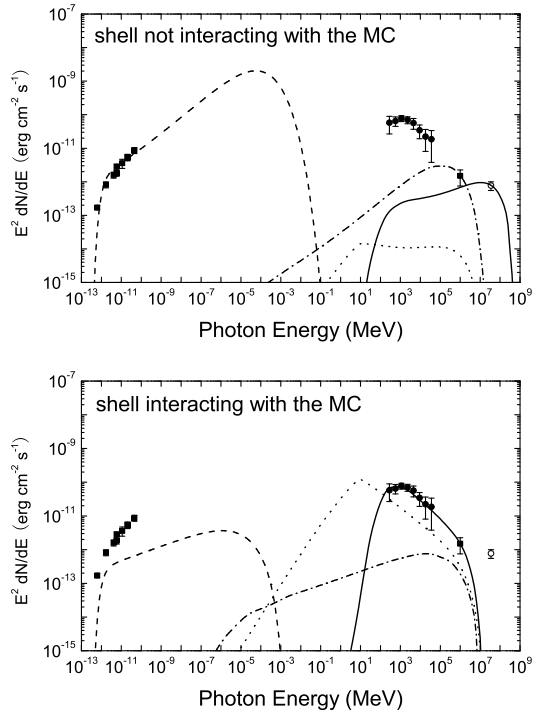}
\caption{\label{figshell1} The resulting synchrotron radiation
(dashed line), bremsstrahlung (dotted line), inverse Compton
scattering (dash-dotted line) and p-p interaction (solid line) for
the shell expanding into the low-density interstellar medium (upper
panel) and for the shell interacting with the MC (lower panel) with
$f=0.4$. The observations in radio \citep[][]{MK94} and high-energy
$\gamma$-rays \citep[][]{Abea09a,Abea09b,Fiea09}. The other
parameters for the upper panel are $d = 6$ kpc, $R_{\rm snr}=30$ pc,
$T_{\rm up}=0.5\times10^6$ K, $M_0=10$, $n_{\rm ism}=0.08$
cm$^{-3}$, $K_{ep}= 2\times10^{-2}$, $B=150$ $\mu$G. Those for the
lower panel are $M_0=2.8$, $n_{\rm ism}=21.4$ cm$^{-3}$, $B=15$
$\mu$G, and the others are the same as the upper panel. In the lower
panel, $n_{\rm ism}$ is determined by making the resulting flux at 9
GeV consist with the one observed by the {\it Fermi} LAT, and the
error of $n_{\rm ism}$ is $n_{\rm ism}=6.0$ cm$^{-3}$ due to the
uncertainty in the observation.}
\end{figure}

Observations of the OH masers \citep[][]{Gea97} and the shocked
atomic and molecular gas \citep[][]{KM97a,KM97b} in the direction of
W51C indicate that a part of the SNR shell is interacting with the
MC. As a result, the velocity of this part of the shell decreases
significantly, and the observed $\gamma$-rays can be reproduced as
p-p collisions accelerated in this part of the shell due to its low
Mach number. The other part of the shell can freely expand into the
interstellar medium with a relatively low density. Therefore, we
assume about $f=0.4$ of the SNR shell transports into the MC with a
Mach number of $2.8$ now, and the other part of the shell have a
relatively large Mach number, $M_0=10$, corresponding to a shock
speed of 830 km s$^{-1}$ for $T_{\rm up}=0.5\times10^6$ K. The
synchrotron photons in the lower energy band are attenuated via
free-free absorption in the journey from the source to the earth,
and the absorption coefficient can be described as \citep[][]{GS65}
\begin{equation}
\label{aff} \alpha_{\rm ff}(\nu)=10^{-2}\frac{n_e^2}{\nu^2
T_e^{3/2}}\left [17.7 + \ln\left(\frac{T_e^{3/2}}{\nu} \right)
\right],
\end{equation}
where $\nu$ is the electron frequency, $n_e$ and $T_e$ are the
density and temperature of the intervening ISM, respectively. We use
$T_e=8000$ K and $n_e d =9.3\times10^{22}$ cm$^{-2}$ for W51C. The
resulting multiband nonthermal emission for the two parts of the
shell are shown in Fig.\ref{figshell1}. The soft photons involved in
the inverse Compton scattering include the cosmic microwave
background, infrared background light and optical star light, with
the energy densities for the three component 0.25, 0.9 and 1.0 MeV
cm$^{-3}$, respectively, similar to those in \citet[][]{Abea09b}.


The observed emission with a shell-type morphology in the radio band
from W51C can be explained as the synchrotron radiation from the
electrons in the shell not encountered by the MC with a Mach number
of 10. A magnetic field strength $>100$ $\mu$G is usually derived
for SNRs \citep[e.g.,][]{Pea06,Taea08}. With $B=150$ $\mu$G, the
maximum energy of the protons is $\sim 1.6\times10^3$ TeV
(Eq.\ref{eq:Pmax}) for an age of $3\times10^4$ yr, which is near the
``knee" energy of the Galactic cosmic rays. With a density of
$n_{\rm ism}=0.08$ cm$^{-3}$, the resulting flux from p-p
interactions at 35 TeV is consistent with the observation with
Milagro.  The electron to proton ratio $K_{ep}$ cannot be
theoretically determined now, which is usually treated as a
parameters limited by the comparison of the resulting flux from the
model with the observations for a given source
\citep[e.g.,][]{Bet02,Mea09,Fea08,Fea09,ZA10}. Its value can be from
$\sim10^{-4}$ for young SNRs to $\sim10^{-2}$ for old ones
\citep[][]{ZA10}. In our model with $M_0=10$ and $B=150$ $\mu$G for
the part of the shell of W51C not colliding with the MC, the radio
must be $2\times10^{-2}$ to have the detected radio emission
explained as the synchrotron radiation of the electrons, which is
about twice the value in the galactic cosmic rays at 10 GeV around
the earth. In order to produce the observed fluxes in the radio
bands, the ratio $K_{ep}$ cannot be very smaller than 0.02 except a
much stronger magnetic field is used. In such a case, the emission
above 1 TeV is dominated by p-p interactions of the accelerated
protons containing a kinetic energy of $3.4\times10^{50}$ erg, and
the TeV photons at 35 TeV observed with Milagro are from this
mechanism (see the upper panel in Fig.\ref{figshell1}).  The maximum
energy of the accelerated electrons is 1.7 TeV (Eq.\ref{eq:Emaxe}).
With the free-free absorption taken into account, the detected
fluxes in the radio band can be reproduced with the synchrotron
radiation of the accelerated electrons.

For the part of the SNR shell interacting with the MC, with
$M_0=2.8$, $n_{\rm ism}=21.4$ cm$^{-3}$ is needed to make the
resulting flux of p-p collisions consist with that at 9 GeV detected
with the {\it Fermi} LAT. The flux from the accelerated electrons is
sensitive to the electron to proton ratio $K_{ep}$, which is set to
$2\times10^{-2}$ as for the part not interacting with the MC. With
this ratio, the magnetic field downstream of the shock cannot be
much larger than $15$ $\mu$G to avoid the resulting synchrotron flux
to conflict with the observed flux points in the radio band. In such
a case, the maximum energies of the electrons ($E_{{\rm max}, e}$)
and the protons ($E_{{\rm max}}$) are 3.0 and 12 TeV, respectively.
With these reasonable parameters, the high-energy $\gamma$-ray
spectrum of the SNR W51C observed with the {\it Fermi} LAT and
H.E.S.S. can be well reproduced (see the lower panel in
Fig.\ref{figshell1}). These $\gamma$-rays are mainly produced via
p-p interaction between the protons in the shell interacting with
the MC and the ambient matter, and the kinetic energy contained in
the protons is about $5\times10^{50}$ erg.

There are uncertainties on the parameters we used to reproduce the
multiband observed flux points for the SNR W51C in our model. For
instance, the value of the ambient density is sensitive to the Mach
number and the whole number of the accelerated particles, which is
determined by the emitting volume with a known spectra of the
accelerated protons. In this paper, the ratio of the shell colliding
with the ambient MC is assumed to be $f=0.4$ based on the observed
extension of the $\gamma$-rays is similar as that in the radio
continuum. For the part of the shell interacting with the remnant, a
Mach number of 2.8 is derived to make the resulting fluxes from p-p
collisions consist with the shape of the flux obtained with the {\it
Fermi} LAT. With such a Mach number, a density of $n_{\rm ism}=21.4$
cm$^{-3}$ is needed to reproduced the observed flux in the
$\gamma$-ray band. The distribution of the protons is $\propto
n_{\rm ism}$, and this relation is also valid for the rate of p-p
interaction. As a result, we can only limit the value $f n_{\rm
ism}^2$ from equaling the resulting flux of p-p interactions to the
observed one, and the density varies from 25 cm$^{-3}$ for $f=0.3$
to 19 cm$^{-3}$ for $f=0.5$. Moreover, the error of the observed
$\gamma$-ray fluxes is about 50\%, and then the uncertainty of
$n_{\rm ism}$ limited in our model is about 6.0 cm$^{-3}$ for
$f=0.4$.

\section{Discussions and summary}
\label{sec:discussion}

W51C is an interesting source for the interaction of the shell with
the MC from the observations of OH masers and shocked atomic and
molecular gases in the direction of the source. It is also the first
SNR detected with the {\it Fermi} LAT in the GeV band. The SNR
appears a shell-type morphology, whereas the GeV signals are clumpy
and extended within the radio shell. The $\gamma$-rays steep above
several GeV, and the spectrum cannot be reproduced via p-p
collisions with an index of $\sim2.0$ for the relativistic protons.
Cosmic ray electrons with a power-law spectrum can be easily
steepened by unity duo to losses \citep[][]{Ka62}, and it is
plausible that this steepening can be explained in the scenario of a
power-law electrons encountering significant losses. However, the
multiband nonthermal radiative properties of W51C have also been
studied by \citet{Abea09b} using a broken power-law spectrum for the
energy distribution of the electrons/protons. They found the
observed flux points in the radio and GeV bands cannot be well
reproduced by assuming the radio emission and high-energy
$\gamma$-rays are produced by the same population of electrons, and
only in the pion-decay scenario in which the $\gamma$-rays are
produced by the protons can the multiband spectrum be well fitted.
Moreover, it seems unlikely that the GeV signals are from the PWN
powered by the putative pulsar \citep{Kea05} because the nebula is
much smaller with a few pc extension in X-rays compared with the GeV
source and its radio emission is much weaker \citep{Abea09b}.

We model the source in the case that one part of the shell
transports into the MC with dense matter and the other part expands
into the interstellar space with a relatively small density. The
spectra of the particles in the two parts of the shell are treated
in a semi-analytical approach to the nonlinear shock acceleration.
Note that in this paper the shock structure is assumed to be
quasi-parallel rather than quasi-perpendicular. The maximum energy
of the protons accelerated in the quasi-perpendicular shocks can be
significantly higher than that in the quasi-parallel cases with a
much smaller perpendicular diffusion efficient compared with the
parallel one, and a quasi-parallel shocks cannot possibly be true
over all of $4 \pi$ \citep[][]{J87}. The contribution of the
secondary $e^{\pm}$ pairs produced from p-p interactions on the
nonthermal emmission is not taken into account in this paper.
Including it needs a time-dependent model to describe the evolution
history of the SNR and the collision between the part of the shock
and the MC, and this complicated process could bring more
assumptions in the model; moreover, with $K_{ep}=2\times10^{-2}$,
the emission of the secondary pairs is usually insignificant
compared with that from the accelerated electrons. In fact, the
multiband observed spectrum can be reproduced with our model, thus
we do not take into account the emission from the secondary
$e^{\pm}$ pairs produced in p-p collisions. Our results indicate
that the $\gamma$-rays detected with the {\it Fermi} LAT and
H.E.S.S. are mainly from the p-p collisions of the protons in the
shell interacting with the MC, whereas the radio emission are
produced via synchrotron radiation of the electrons accelerated in
the shell not interacting with the MC; moreover, the high-energy
$\gamma$-rays obtained with Milagro at 35 TeV are possibly from p-p
interactions of the protons accelerated in the shell not interacting
with the MC.

A MC can be illuminated by the particles escaped from a nearby SNR,
and high-energy $\gamma$-rays can be produced when these particles
diffuse into the cloud \citep[][]{Gea09,Fuea09}. For example,
\citet[][]{Tea08} interpreted the source MAGIC J0616+225 as TeV
emission from the particles escaped from the SNR IC 443 and
diffusing into a MC located about 20 pc from the SNR. Moreover, the
nonthermal radiative properties of the the open cluster Westerlund 2
and the old-age mixed-morphology SNR W28 can also be modeled as MCs
illuminated by high-energy particles escaped from SNRs
\citep[][]{Fuea09}. On the other hand, $\gamma$-rays can also be
effectively produced from a shock wave colliding with a MC
\citep[e.g.,][]{Aea94,ZF08}. $\gamma$-ray emission produced from a
MC overtaken by a shock wave with a low Mach number has also been
used to discuss the emission properties of the TeV sources, HESS
J1745-303 (A) and HESS J1714-385 \citep[][]{FZ08}. In this paper, we
model the SNR W51C as the MC around the remnant overtaken by part of
the shell, and the other part of the shell expands into the
relatively tenuous interstellar medium. The spectrum obtained with
the {\it Fermi} LAT can be reproduced as $\gamma$-rays from p-p
collisions of the protons in the shell with a low Mach number.

\section*{Acknowledgments}
We are very grateful to the anonymous referee for his/her helpful
comments. This work is partially supported by the National Natural
Science Foundation of China (NSFC 10778702, 10803005), a 973 Program
(2009CB824800), and Yunnan Province under a grant 2009 OC.

\label{lastpage}

\end{document}